\documentclass[12pt,a4paper]{article}
\usepackage{amsmath,amssymb,graphicx}

\newcommand{\sdot}{\!\cdot\!}
\newcommand{\SLASH}[1]{/\!\!\! #1}

\title{\bf Dynamical relativistic corrections to the leptonic decay 
width of heavy quarkonia}
\author{F.~Bissey$^{*}$, J.-J.~Dugne$^{\dag}$, J.-F.~Mathiot$^{\#}$}

\date{}

\begin{document}
\maketitle

\begin{center}

{\small \em  Laboratoire de Physique Corpusculaire,
Universit\'e Blaise-Pascal,\\ CNRS/IN2P3, 24 avenue des
Landais, F-63177 Aubi\`ere Cedex, France}\\

\vspace{0.2 cm}
\end{center}

\begin{abstract}
We calculate the dynamical relativistic corrections, originating from 
radiative one-gluon-exchange, to the leptonic decay width of heavy quarkonia 
in the framework of a covariant formulation of Light-Front Dynamics. 
Comparison with the non-relativistic calculations of the leptonic decay width 
of $J\!\!=\!1$ charmonium and bottomonium $S$-ground states shows that 
relativistic corrections are large. Most importantly, the calculation of 
these dynamical relativistic corrections legitimate a perturbative expansion 
in $\alpha_s$, even in the charmonium sector. This is in contrast with the 
ongoing belief based on calculations in the non-relativistic limit. 
Consequences for the ability of several phenomenological potential to 
describe these decays are drawn.

\end{abstract}

\begin{figure}[b]

\noindent  PACS numbers: 03.65.pm,13.20.Gd,12.39.Ki\\
\noindent  Keywords: Heavy quarkonium, Light-Front Dynamics, leptonic
decays \\
\noindent  PCCF RI 01-16\\

------------------------------------------ \\
$^*$){\small e-mail: bissey@in2p3.fr}\\
$^\dag$){\small e-mail: dugne@clermont.in2p3.fr}\\
$^\#$){\small e-mail: mathiot@in2p3.fr}
\end{figure}

\newpage

\section{Introduction}

Although the structure of heavy quarkonia in terms of a heavy quark-antiquark 
non-relativistic bound state is known for a long time, many recent 
developments show the importance of a relativistic description of such 
states. Mainly two paths are followed:
\begin{enumerate}  
\item One can first estimate relativistic corrections to the $\bar{q}q$ 
bound state in a $v/c$ expansion. This is the aim of the NRQCD 
formalism \cite{NRQCD}. When the quark mass is large enough, this is 
certainly adequate since this expansion should converge rapidly. It is 
however not clear whether the charm quark mass is heavy enough for such an 
expansion to be valid. To check it out, one should compare the results of 
NRQCD with relativistic  calculations.  
\item Relativistic corrections can also be calculated in a
relativistic framework to describe the two-body bound state, such as
the Bethe-Salpeter formalism, or Light-Front Dynamics (LFD). In a previous 
article \cite {loui00}, we have investigated the relativistic kinematical 
corrections to the leptonic decay width of the $J/\psi$ and the $\Upsilon$ 
induced by the finite momentum between their quark and antiquark. We used an 
appropriate formalism, the so called "Covariant 
formulation of Light-Front Dynamics" (CLFD), the details of which can be found
in Ref.~\cite{carb98}.  As pointed out in this reference, the CLFD allows a 
straightforward comparison between relativistic and non-relativistic 
calculations.  
\end{enumerate}
We shall investigate in this article the importance of dynamical relativistic 
corrections. Kinematical relativistic corrections can be calculated without 
any knowledge of the dynamical origin of the two-body wave function. 
Dynamical corrections, on the other hand, correspond to relativistic
corrections to the wave function itself.  While the non-relativistic two-body 
wave function of the $J=1$ state has only two components (the $S$ and $D$ 
states), the relativistic wave function formulated on the light-front has 6 
dynamical components \cite{carb98}. If the dynamical origin of the two-body 
interaction is known, these components can, in principle, be calculated 
exactly or else calculated in perturbation theory, starting from the 
non-relativistic components. This has been done for instance in 
Ref.~\cite{ck-deut}, for the calculation of the relativistic corrections to 
the deuteron wave function.

In the case of heavy quarkonium, the two body interaction has two main parts: 
a confining interaction, and the interaction coming from the exchange of an 
effective gluon (One-Gluon-Exchange, OGE). The latter interaction gives rise, 
in the non-relativistic limit, to the well known Coulomb interaction. While 
little is known on the dynamical origin of the confining potential, one may 
hope that for heavy quarkonia, the two-body wave function is mostly sensitive, 
for relativistic corrections, to the short range part of the interaction,
{\it i.e.} to the one-gluon exchange potential and its $1/r$ non-relativistic 
behaviour.

This is the usual assumption made when calculating the dynamical relativistic 
correction coming from the OGE interaction. In zeroth order in a 
$\vec{k}^2/m^2$ expansion, where $\vec{k}$ is the relative momentum of the 
quark-antiquark pair in the two-body bound state, it gives rise to the 
following well known correction to the leptonic decay width of $J=1$ 
states \cite{barb76,celm79}:
\begin{equation}
\label{gg}
\Gamma_1^{NR}= 
 \Gamma_0^{NR} \left( 1-\frac{16\alpha_s}{3\pi} \right)\text{ .}  
\end{equation}
In the case of the $J/\Psi$ leptonic decay width, this contribution amounts 
to a $50\%$ reduction. This large correction clearly calls for a 
relativistic calculation of this contribution to all orders in 
$\vec{k}^2/m^2$. This is the aim of the present article.  

We shall follow very closely the procedure of  Ref.~\cite{loui00}. We refer 
the reader to this reference, and to Ref.~\cite{carb98} for a review article 
on CLFD. We introduce in section 2 the scheme of the calculation of the
leptonic decay width. The dynamical relativistic corrections are introduced 
in section 3.  Numerical results are presented in section 4, and our 
conclusions and perspectives are drawn out in section 5.

\section{The leptonic decay width of heavy quarkonium}

\subsection{The quarkonium wave function in CLFD}

Among the various ways to deal with relativity in the description of bound 
states, we will focus  on LFD. In the standard formulation of LFD \cite{bpp}, 
the wave function of a bound system is defined on a plane characterised by the 
equation $t+z/c=0$. The usual Schr\"odinger, equal time, formalism is easily 
recovered in LFD by letting $c$ go to infinity.

The description of relativistic heavy quark systems in LFD has many 
advantages, of which the most important one may be the absence of vacuum 
fluctuations. Accordingly, a meaningful decomposition of the state vector 
describing the system in terms of Fock components of definite number of 
particles is possible.  Of course, the number of Fock components to be 
considered in any practical calculation depends on the dynamics of the 
system, and on the kinematical regime one is interested in.

The most serious drawback of this formulation, however, is that the position 
of the light-front $t+z=0$ (with $c=1$) is not rotational invariant. Since 
rotations in the $zx$ and $zy$ planes change the position of the light-front, 
the associated generators shall depend on the dynamics and cannot be reduced 
to kinematical transformations \cite{dirac}. This means that one needs to 
know the complete dynamics in order to write down the general structure of a
bound state of definite angular momentum.  It also means that any 
electromagnetic operator should have the same (dynamical) transformation 
properties to match the bound state wave function one's. This is essential to 
guarantee that any physical amplitude (or cross-section) does not depend on 
the particular choice of the light-front we start with.

We therefore need an explicit procedure to exhibit in a convenient way these 
dynamical transformations.  This is  achieved in the covariant formulation of 
LFD. Our starting point is the invariant definition of the light-front by 
$\omega \sdot x=0$ where $\omega$ is a (unspecified) light-like four vector 
($\omega ^2=0$). The standard formulation of LFD can be easily recovered with 
the particular choice $\omega=(1,0,0,-1)$ for the light-like four vector.

This definition of the light-front is explicitly invariant by any 
four-dimensional rotations, or any three-dimensional rotations and Lorentz 
boosts. Consequently, these transformations become $\omega$-dependent, but do 
not necessitate the knowledge of the dynamics of the system to construct them 
explicitly. The dynamics enters now into the $\omega$-dependence of the 
wave function and of the electromagnetic operator, in such a way that any 
physical amplitude should not depend on the particular position of the 
light-front, {\it i.e.} on $\omega$, unless approximations have been made. In 
this case, which is almost always true in practice, the explicit covariance 
of the approach enables us to exhibit the $\omega$-dependence of the 
amplitude and to separate the physical part from the non-physical, 
$\omega$-dependent one, as we shall explain below for the leptonic decay 
amplitude.

The wave function ${\Phi}$, of a two-body bound state, can be decomposed in 
terms of all the possible independent spin structures compatible with the 
quantum numbers of the state studied. In the particular case of vector 
mesons, which we are interested in, we can write :
\begin{equation}
\label{eq:2}                                                
\Phi^\lambda_{\sigma_2\sigma_1}(k_1,k_2,p,\omega)=
\sqrt{m} e^{\lambda}_{\mu}(p) 
\bar{u}^{\sigma_2}(k_2)\phi^\mu v^{\sigma_1}(k_1)\text{ ,}        
\end{equation}                      
with:
\begin{multline}
\label{nz2} 
\phi^{\mu}=
\varphi_1\frac{(k_1 - k_2)^\mu}{2m^2}+\varphi_2\frac{1}{m}\gamma^\mu   
+\varphi_3\frac{\omega^\mu}{\omega\sdot p} 
+\varphi_4\frac{(k_1-k_2)^\mu\SLASH{\omega}}{2m\omega\sdot p} \\             
+\varphi_5\frac{i}{2m^2\omega\sdot p}\gamma_5 
\epsilon^{\mu\nu\rho\gamma}(k_1+k_2)_\nu (k_1-k_2)_\rho \omega_\gamma
+\varphi_6\frac{m\omega^\mu \SLASH{\omega}}{(\omega\sdot p)^2}\text{ .}      
\end{multline}
This decomposition is similar to the decomposition of the deuteron wave 
function found in Ref.~\cite{ck-deut}. The six components of the wave 
function, $\varphi_1$-$\varphi_6$, depend on two invariants. In order to 
make a close connection to the non-relativistic case, it will be convenient 
to introduce another pair of variables \cite{carb98} defined by:
\begin{equation}
\label{sc4}                                                     
\vec{k}=L^{-1}({\cal P})\vec{k}_1 = 
\vec{k}_1 -\frac{\vec{\cal P}}{\sqrt{{\cal P}^2}}
\left[k_{10}-
 \frac{\vec{k}_1\sdot \vec{{\cal P}}}{\sqrt{{\cal P}^2}+{\cal P}_0}
\right]
\text{ ,}              
\end{equation}
\begin{equation}
\label{sc5}                                                     
\vec{n}=
\frac{L^{-1}({\cal P})\vec{\omega}}{\vert L^{-1}({\cal P})\vec{\omega}\vert} 
=\sqrt{{\cal P}^2}\frac{L^{-1}({\cal P})\vec{\omega}}{\omega\sdot p}\text{ ,}
\end{equation}                                                                 
where:                                                                          
\begin{equation}
\label{calp}                                                    
{\cal P} = k_1+k_2\text{ .} 
\end{equation}
The relativistic momentum $\vec{k}$ corresponds, in the frame where 
$\vec{k}_1+\vec{k}_2=\vec{0}$, to the usual relative momentum between the two 
particles. Note that this choice of variable does not assume that we restrict 
ourselves to this particular frame. The unit vector $\vec{n}$ corresponds, in 
this frame, to the spatial direction of $\vec{\omega}$. In terms of these
variables, the wave function takes a form similar to the non-relativistic one:
\begin{equation}
\label{nz7}
\Psi^\lambda_{\sigma_2\sigma_1}(\vec{k},\vec{n}) = 
\sqrt{m}w^\dagger_{\sigma_2} 
\psi^\lambda (\vec{k},\vec{n})w_{\sigma_1}\text{ ,} 
\end{equation} 
with:
\begin{multline}
\label{nz8} 
\vec{\psi}(\vec{k},\vec{n}) = 
f_1\frac{1}{\sqrt{2}}\vec{\sigma} 
+f_2\frac{1}{2}
\left(\frac{3\vec{k}(\vec{k}\sdot\vec{\sigma})}{\vec{k}^2}-\vec{\sigma}\right)
+f_3\frac{1}{2}
\left(3\vec{n}(\vec{n}\sdot\vec{\sigma})-\vec{\sigma}\right)\\  
+f_4\frac{1}{2k}\left(3\vec{k}(\vec{n}\sdot\vec{\sigma})+ 
3\vec{n}(\vec{k}\sdot\vec{\sigma})-
2(\vec{k}\sdot\vec{n})\vec{\sigma}\right)         
+f_5\sqrt{\frac{3}{2}}\frac{i}{k}\left[\vec{k}\times \vec{n}\right]
+f_6\frac{\sqrt{3}}{2k}
\left[\left[\vec{k}\times \vec{n}\right]\times\vec{\sigma}\right]\text{ ,}
\end{multline}
where $w_\sigma$ is the two-component Pauli spinor normalised to 
$w^\dagger_\sigma w_\sigma=1$, and $\vec{\sigma}$ are the usual Pauli 
matrices.  The relation between $\psi^\lambda$ and $\vec{\psi}$, is the same 
as the relation between the spherical function $Y_1^\lambda(\vec{n})$ and
$\vec{n}$.

The coefficients of the spin structures in Eq.~(\ref{nz2}) and Eq.~(\ref{nz8})
are scalar functions of two independent invariants, which we can choose to be
$\vec{k}^2$ and $\vec{k}\sdot\vec{n}$, since these variables are only
rotated by a Lorentz boost \cite{carb98}. In the non-relativistic limit, only
two components remain, $f_1$ and $f_2$, and they only depend on 
$\vec{k}^2$. This can be easily seen if we keep track of the $c$ factors, 
and then let $c$ goes to infinity to get the non-relativistic limit. In this 
study of heavy quarkonium states, one may neglect the tensor component, 
$f_2$, so one is left with the non-relativistic wave function $f_1\equiv
\phi^{NR}(\vec{k}^2)$. The relation between $\varphi_1$, $\varphi_2$ and 
$\phi^{NR}$ is given by \cite{carb98}:
\begin{eqnarray}        
\label{eq:9}
\varphi_1(\vec{k}^2) & = & 
\frac{m^2}{4\varepsilon_k (\varepsilon_k +m)}
\sqrt{2}\phi^{NR}(\vec{k}^2)\text{ ,}\\
\label{eq:10}
\varphi_2(\vec{k}^2) & = & 
\frac{m}{4\varepsilon_k} \sqrt{2}\phi^{NR}(\vec{k}^2)\text{ ,}
\end{eqnarray}
where $\varepsilon_k=\sqrt{\vec{k}^2+m^2}$. Note that the wave function 
$f_1\equiv \phi^{NR}$ is normalised according to: 
\begin{equation} 
\label{norf}
\int \left\vert\phi^{NR}(\vec{k}^2)\right\vert^2 
\frac{d^3k}{(2\pi)^3}\frac{m}{\varepsilon_k}=1\text{ .}
\end{equation}

\subsection{The leptonic decay width} 

Since the total physical amplitude for the process $M_J \to e^+ e^-$ can 
be factorised into two separate amplitudes $M_J \to \gamma$ and 
$\gamma \to e^+ e^-$, the relevant physical information is completely 
included in the amplitude ${\cal M}^{\mu \rho}$ to produce a photon with 
polarisation $\epsilon^\mu$ from a vector quarkonium state of 
polarisation $\epsilon^\rho$.

Since our formulation of LFD is explicitly covariant, we can decompose 
${\cal M}^{\mu\rho}$ in terms of all possible tensor structures build up with 
the two four-momenta at our disposal, $p$ and $\omega$, where $p$ is the 
four-momentum of the quarkonium \cite{loui00}. Thus, we can  write:
\begin{equation}
{\cal M}^{\mu\rho} = F a_1^{\mu\rho} +
\frac{B_{1}}{2\omega \sdot p} a_2^{\mu\rho} + 
\frac{B_{2}}{2\omega \sdot p} a_3^{\mu\rho} + 
B_{3}\frac{M^2}{(\omega \sdot p)^2} a_4^{\mu\rho} + 
D a_5^{\mu\rho}\text{ ,} 
\label{E7}
\end{equation}
with:
\begin{subequations}
\label{eq19}
\begin{eqnarray}
a_1^{\mu\rho} & = & g^{\mu\rho}-\frac{p^\mu p^\rho}{M^2}\text{ ,}\\
a_2^{\mu\rho} & = & p^\mu \omega^\rho + p^\rho \omega^\mu \text{ ,}\\
a_3^{\mu\rho} & = & p^\mu \omega^\rho - p^\rho \omega^\mu \text{ ,}\\
a_4^{\mu\rho} & = & \omega^\mu \omega^\rho \text{ ,}\\
a_5^{\mu\rho} & = & \frac{p^\mu p^\rho}{M^2}\text{ .}
\end{eqnarray}
\end{subequations}
We have denoted by $M$ the mass of the quarkonium. Note that the term 
proportional to $D$  does not contribute to the leptonic decay width. The 
amplitude ${\cal M}^{\mu\rho}$ has two kinds of terms. The first one, 
proportional to $F$, is the physical contribution to the decay width. The 
other three, proportional to $B_{1}$, $B_{2}$ and $B_{3}$ are 
$\omega$-dependent contributions. In an exact calculation, the coefficients 
$B_{1}$, $B_{2}$ and $B_{3}$ should be zero since the physical leptonic decay 
width should not depend on the particular orientation of the light-front one 
starts with. However, in any approximate  calculation, these terms may be 
non-zero, but they are not physical. Consequently, they have to  be 
eliminated in the computation of the physical leptonic width.

To extract $F$ from the general amplitude ${\cal M}^{\mu\rho}$, one
can first multiply ${\cal M}^{\mu\rho}$ successively by the five tensor 
structures $a_{1}$ to $a_{5}$ given in Eqs.~(\ref{eq19}). This gives a system 
of five coupled equations which is solved to get the physical amplitude $F$.  
Thus, we find:
\begin{equation}
F=\frac{1}{2} \left(I_1-2I_2 +I_4 +I_5 \right)\text{ ,}
\label{E8}
\end{equation}
with:
\begin{equation}
I_i={\cal M}_{\mu\rho} a_i^{\mu\rho}\text{ .}
\label{E9}
\end{equation}
These quantities are easily evaluated once ${\cal M}^{\mu \rho}$ is
calculated from the process under consideration, using the diagrammatical
rules given in Ref.~\cite{carb98}. Given $F$, the decay width can be 
calculated \cite{loui00}, and is given by:
\begin{equation} 
\label{gamma}
\Gamma=\frac{4 \pi}{3 M^3}\alpha^2 e_q^2 \vert F\vert^2\text{ ,}
\end{equation}
where $e_q$ is the electric charge of the quark and $\alpha$ is the 
electromagnetic fine structure constant.

\subsection{The zeroth order calculation} 
\label{S:2:3}
\begin{figure}
\centering\includegraphics[clip=true]{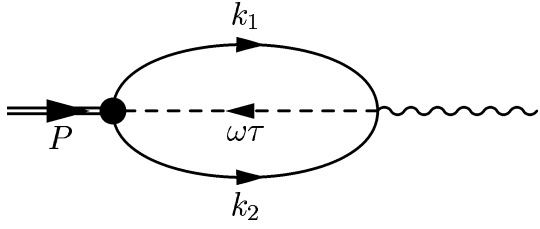}
\caption{Leading order contribution to the leptonic decay width. The dashed 
line represents the spurion (see text for details).}
\label{Fig:1} 
\end{figure}

The leading order contribution to the leptonic decay is shown in 
Fig.~\ref{Fig:1}, where we have removed for simplicity the trivial vertex 
$\gamma \to e^+ e^-$. Off-energy shell effects are governed by the variable 
$\tau$, which is unambiguously determined by four-momentum conservation and 
the on mass-shell condition for each particle \cite{carb98}. Using the
conservation law at the bound state vertex, we have\footnote{One should note 
that in CLFD, in the frame where we have $\vec{k}_1 +\vec{k}_2 =\vec{0}$,
the total momentum $\vec{p}$ is not $\vec{0}$. Instead 
$\vec{p}+\vec{\omega}\tau$ is $\vec{0}$.}:
\begin{equation}
\label{eq:17}
k_1+k_2 = p+\omega\tau \text{ .}   
\end{equation}
To keep track of this 
conservation law, we represent the four-vector $\omega\tau$ by a dashed line 
in every diagram (the so-called spurion line, see Ref.~\cite{carb98} for 
more details). Note that the outgoing photon has virtuality $q^2=p^2$. 
However, it can be assumed to be a "physical particle" in the sense that the 
final process $\gamma \to e^+ e^-$ is completely disconnected from the decay 
process $M_J \to \gamma$, and is exactly calculable in terms of outgoing free 
electron and positron.
As already explained in details in Ref.~\cite{loui00}, the zeroth order 
contribution to the amplitude, represented in Fig.~\ref{Fig:1}, is given by:
\begin{equation}
\label{eq22}
{\cal M}^{\mu\rho}=
-\sqrt{3m}\int \frac{d^3k_1}{(2\pi)^{3}\,2\varepsilon_{k_1}(1-x)}
Tr\left[\gamma^\mu(\SLASH{k}_1+m)\phi^\rho (m-\SLASH{k}_2)\right]\text{ ,} 
\end{equation}
where $x=\omega\sdot k_1/\omega\sdot p$ and $\phi^\rho$ is defined in
Eq.~(\ref{nz2}). The physical part of the amplitude in zeroth order
calculated by Eq.~(\ref{E8}), $F_0$, can now be written in the form:
\begin{equation} 
\label{f1}
F_0=\int \frac{d^3 k}{(2\pi)^3} \frac{m}{\varepsilon_k} O_0
(\vec{k}^2)\phi^{NR}(\vec{k}^2)\text{ ,}
\end{equation}
where $O_0(\vec{k}^2)$ is:
\begin{equation}
\label{eq:20}
O_0(\vec{k}^2)=
-2\sqrt{6m}\left[1+\frac{2}{3}\frac{m}{\varepsilon_k}
\left(1-\frac{\varepsilon_k}{m}\right)^2\right]\text{ .}
\end{equation}
Note that in leading order in a $\vec{k}^2/m^2$ expansion, kinematical
relativistic corrections originate only from the factor $m/\varepsilon_k$ of 
the relativistic phase space volume. In the non-relativistic limit, one has:
\begin{equation}
\label{eq:21}
F_0^{NR}=-2\sqrt{6m}\phi^{NR}(r=0)\text{ .}
\end{equation}

\section{Dynamical relativistic corrections}

\subsection{Radiative corrections} 
\label{S:3:1}

\begin{figure}
\centering\includegraphics[clip=true]{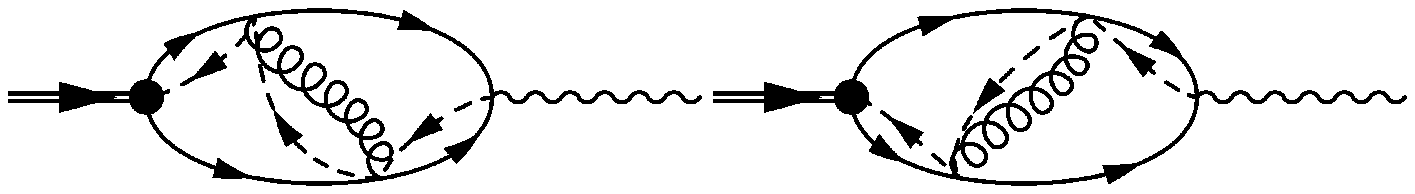}
\caption{Radiative correction to the leptonic decay width.}
\label{fig2}

\centering\includegraphics[clip=true]{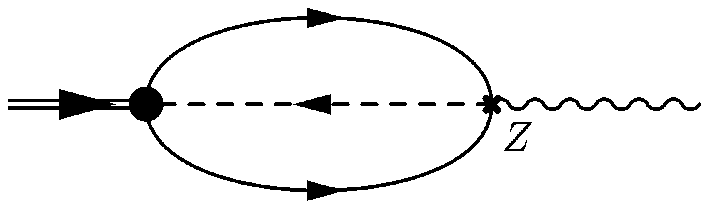}
\caption{Vertex renormalisation contribution to the leptonic decay width.}
\label{fig3}
\end{figure}

As already pointed out, the dynamical relativistic corrections necessitate 
the knowledge of the dynamical origin of the two-body wave function, 
{\it i.e.}, the way quark-antiquark states are bound. This goal is still far 
from being achieved. The standard assumption to overcome this issue is to 
suppose that for quark masses heavy enough, the dynamics is governed by a 
perturbative one-gluon-exchange. The physical amplitude ${\cal M}^{\mu\rho}$ 
for the relevant processes are indicated in Fig.~\ref{fig2}. They can be 
calculated analogously to Eq.~(\ref{eq22}) using the diagrammatical rules 
given in Ref.~\cite{carb98}. The details of the calculation are too lengthy 
to be shown here, but present no particular difficulties. The amplitude is 
both ultraviolet and infrared divergent. The computation of these 
contributions is detailed in appendix. 

Since these diagrams are ultraviolet divergent, we also
need to include the renormalisation of the quark charge at the 
electromagnetic vertex, as has been already detailed in Ref.~\cite{dugn01}. 
The corresponding diagram for this contribution is shown in Fig.~\ref{fig3}. 
The counter-term $Z$ is given by:
\begin{equation}
\label{Z}
Z=
\frac{4}{3} \alpha_s
\left[\frac{9}{8\pi}+\frac{1}{2\pi}\log\left(\frac{\mu^2}{m^2}\right)
+\frac{1}{4\pi}\log\left(\frac{\Lambda^2}{m^2}\right)\right]\text{ ,}
\end{equation}
where the leading $\frac{4}{3}$ is a factor of colour generated by the gluon
exchange.

The ultraviolet regularisation of the amplitude is done according to the 
Pauli-Villars prescription, with a cut-off $\Lambda$. This is the same 
prescription as in Ref.~\cite{dugn01}. The amplitude is made infrared finite 
by giving a small non zero mass $\mu$ to the photon. We shall see in the next 
section that the final contribution remains infrared finite, as it should.

\subsection{Coulomb subtraction}

As it is well known, one should be very careful in the evaluation of
dynamical relativistic corrections as indicated in Fig.~\ref{fig2}. Indeed, 
when we evaluate the zeroth order contribution shown in Fig.~\ref{Fig:1}, with 
the non-relativistic wave function, $\phi^{NR}(\vec{k}^2)$, solution of the 
Schr\"odinger equation with the non-relativistic two-body potential:
\begin{equation} 
\label{c1}
V_{\bar{q} q}(r)=V_{OGE}(r)+V_{Conf}(r)\text{ ,}
\end{equation}
we implicitly incorporate corrections of the type shown in Fig.~\ref{fig2}, in 
leading non-relativistic order at least. One should therefore make sure that 
these contributions are properly removed from the relativistic calculation in 
order to avoid double counting.

The usual procedure to do this is to remove from the relativistic OGE the 
Coulomb interaction. We shall see below that our formulation of the leptonic 
decay width in CLFD enables us to have a clear handle on the contribution to 
remove, at any level of approximation.

\begin{figure}
\centering\includegraphics[clip=true]{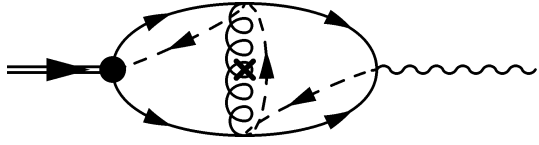}
\caption{Coulomb contribution to the radiative correction of the leptonic 
decay width.}
\label{fig4}
\end{figure}

Let us start from the Schr\"odinger equation, written in the form:
\begin{equation} 
\label{c2}
\phi^{NR}(\vec{k}^2)=
-\frac{4m}{s-M^2}\int \frac{d^3 k'}{(2\pi)^3} 
V_{\bar{q}q}(\vec{k'}^2)\phi^{NR}(\vec{k'}^2)\text{ ,}
\end{equation}
with $V_{\bar{q}q}$ given by Eq.~(\ref{c1}), and $s=4(\vec{k}^2+m^2)$. The 
contribution to $\phi^{NR}(\vec{k}^2)$ which corresponds in perturbation 
theory (first order in $\alpha_s$) to the OGE interaction in the 
non-relativistic limit is therefore given by:
\begin{equation} 
\label{c3}
\delta \phi^{NR}(\vec{k}^2)=
\frac{4}{3} \frac{4m}{s-M^2}\int \frac{d^3 k'}{(2\pi)^3} 
\frac{\alpha_s}{(\vec{k}-\vec{k'})^2}\phi^{NR}(\vec{k'}^2)\text{ .}
\end{equation}
Inserted in the zeroth order calculation indicated in Fig.~\ref{Fig:1}, 
and written in Eq.~(\ref{f1}), it gives the contribution, $\delta F$, to the 
physical amplitude:
\begin{equation} 
\label{c4}
\delta F=
\int \frac{d^3 k}{(2\pi)^3} \frac{m}{\varepsilon_k} 
O_0(\vec{k}^2)\delta \phi^{NR}(\vec{k}^2)\text{ .}
\end{equation}
This contribution is represented graphically in Fig.~\ref{fig4}. It is 
ultraviolet finite, but diverges in the infrared region. We regularise the 
latter divergence by giving a small finite mass $\mu$ to the gluon. Note also 
that in Eq.~(\ref{c3}), the phase space volume in $\vec{k'}$ is the
non-relativistic one since it involves the non-relativistic wave function, 
solution of the Schr\"odinger equation. However, in Eq.~(\ref{c4}), it 
involves the relativistic phase space volume since one has to perform the
integral over $\vec{k}$ on the whole momentum range, as done in 
Eq.~(\ref{f1}) for the calculation of kinematical relativistic corrections.

The contribution $\delta F$ of Eq.~(\ref{c4}), together with 
$\delta\phi^{NR}(\vec{k}^2)$ given in Eq.~(\ref{c3}), should be removed from 
the total relativistic amplitude ${\cal M}^{\mu \rho}$ given by 
Eq.~(\ref{eq22}). In the non-relativistic limit for the two-body bound state 
(limit $(\vec{k'}^2/m^2) \ll 1$), it gives the following contribution to $F$:
\begin{equation} 
\label{c5}
\delta F^{NR}=\frac{4}{3}\alpha_s \frac{m}{\mu}F_0^{NR}\text{ .}
\end{equation}

\subsection{Non-relativistic limit}

In the non-relativistic limit for the two-body bound state, the contribution 
of Fig.~\ref{fig2} to the physical amplitude is:
\begin{equation} 
\label{FOGE}
F_{OGE}^{NR}=\frac{4}{3}\alpha_s
\left[\frac{m}{\mu}-\frac{7}
  {8\pi}+\frac{1}{2\pi}\log\left(\frac{\mu^2}{m^2}
  \right)+\frac{1}{4\pi}\log\left(\frac{\Lambda^2}{m^2} 
  \right) \right] F_0^{NR}\text{ ,}
\end{equation}
while the contribution of Fig.~\ref{fig3} gives, from Eq.~(\ref{Z}):
\begin{equation} 
\label{FZ}
F_Z^{NR}=
 -Z\, F_0^{NR}=
 -\frac{4}{3}\alpha_s
 \left[\frac{9}{8\pi}+
 \frac{1}{2\pi}\log\left(\frac{\mu^2}{m^2}\right)+
 \frac{1}{4\pi}\log\left(\frac{\Lambda^2}{m^2} 
 \right) \right] F_0^{NR}\text{ .}
\end{equation}
Adding the contributions of $F_{OGE}^{NR}$ and $F_Z^{NR}$, then removing the 
contribution $\delta F^{NR}$, we get the final, infrared finite, contribution 
in the non-relativistic limit:
\begin{equation}
\label{eq:30}
F_1^{NR}=-\frac{8}{3\pi}\alpha_s F_0^{NR}\text{ .}
\end{equation}
Added to $F_0^{NR}$, it gives the well known radiative correction in leading 
order in $\alpha_{S}$, as indicated in Eq.~(\ref{gg}) for the total width.
        
\section{Numerical results}

To easily compare our numerical calculations with the non-relativistic limit 
found above, we consider the following ratio:
\begin{equation}
\label{eq:31} 
R \equiv \frac{\Gamma_1}{\Gamma_0^{NR}}=
\frac{\Gamma_1}{\Gamma_0}\frac{\Gamma_0}{\Gamma_0^{NR}}\text{ ,}
\end{equation}
where $\Gamma_1$ is the relativistic decay width to first order in $\alpha_s$ 
(calculated in sec.~\ref{S:3:1}), $\Gamma_0$ is the relativistic decay width 
in leading order in $\alpha_s$ (calculated in sec.~\ref{S:2:3}), while 
$\Gamma_0^{NR}$ is the non-relativistic limit in leading order in 
$\alpha_s$ given by the Van Royen-Weisskopf formula:
\begin{equation}
\label{g0}
\Gamma_0^{NR}= 16 \pi e_q^2\frac{\alpha^2}{M^2} 
\vert \phi^{NR}(r=0)\vert^2\text{ .}
\end{equation}
We also note:
\begin{eqnarray}
\label{g1}
\frac{\Gamma_0}{\Gamma_0^{NR}} & = & R_K\text{ ,}\\ 
\label{g2}
\frac{\Gamma_1}{\Gamma_0} & = & \left(1-\frac{16 \alpha_s}{3\pi} R_D\right)
\text{ .}
\end{eqnarray}
The ratio $R_K$ is simply the kinematical relativistic corrections we already 
calculated in Ref.~\cite{loui00}. The dynamical corrections have been written 
in the form of Eq.~(\ref{g2}) to explicitly exhibit the non-relativistic 
limit of the radiative corrections when the two-body bound state wave 
function is restricted to very small momenta. In this case, $R_D$ should go 
to unity.

For the hadronic vertex, we use two types of test wave functions to assess 
the importance of these relativistic corrections.  On the one hand, we take 
an harmonic oscillator wave function corresponding to a confining potential 
for large distances between the quark and the anti-quark. The wave function 
reads:
\begin{equation}
\phi_H^{NR}(\vec{k}^2) = N \left(\frac {6 \pi}{k^2_m}\right)^{3/4}
\exp{\left(-\frac {3}{4}\frac{\vec{k}^2}{k^2_m}\right)}\text{ .}
\label{eqn1}
\end{equation}
In the previous expression, $k^2_m$ is the mean relative momentum squared. In 
this case, the wave function at the origin writes:
\begin{equation}
\label{eq:36}
\phi_H^{NR}(r=0) = N \left(\frac {2}{3\pi}\right)^{3/4}
\left(k^2_m\right)^{3/4}\text{ .}
\end{equation}
On the other hand, we take a Coulomb wave function expected to dominate for 
very heavy quark masses. It is given by:
\begin{equation}
\phi_C^{NR}(\vec{k}^2) = 8 N \sqrt{\pi} 
\frac{(k^2_m)^{5/4}}{\left(\vec{k}^2+k^2_m\right)^2}\text{ .}
\label{eqn2}
\end{equation}
As in Eq.~(\ref{eqn1}), $k^2_m$ is the mean relative momentum squared. Thus, 
the wave function at the origin is:
\begin{equation}
\label{eq:38}
\phi_C^{NR}(r=0) = N \frac{1}{\sqrt{\pi}}\left(k^2_m\right)^{3/4}\text{ .}
\end{equation}
Note that the choice of these test wave functions is just a way to 
investigate the sensitivity of the decay constant to various shapes of wave 
function. It is not intended to imply any assumption about the exact dynamics 
binding the heavy quark-antiquark pair. The normalisation factor $N$ is 
calculated according to  Eq.~(\ref{norf}). It goes to one in the 
non-relativistic limit.

\begin{figure} 
\centering\includegraphics[height=20cm,clip=true]{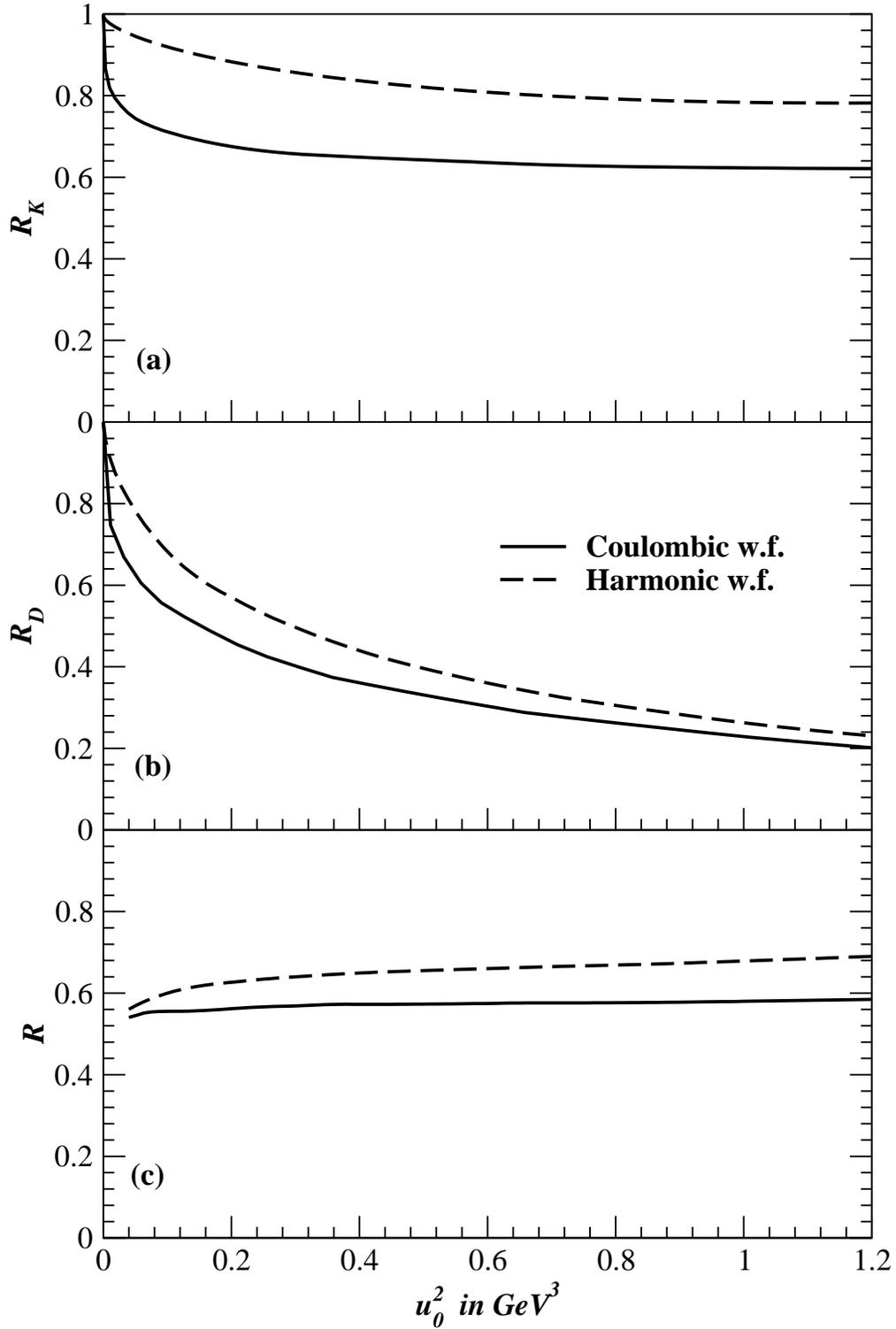}
\caption{Relativistic corrections to the $J/\Psi$ leptonic decay width, for 
the two test wave functions. The Coulombic one (solid line) and the Harmonic 
one (dashed line). These contributions are plotted as a function of $u_0^2$, 
defined in Eq.~(\ref{eq:39}) and proportional to the square of the wave 
function at the origin. Sub-figure (a) gives the kinematical correction, 
ratio $R_K$, (b) give the dynamical correction, ratio $R_D$, and (c) show the 
total correction, ratio $R$. The mass of the charm quark is taken to be 
$M_{\Psi}/2$.}
\label{fig5} 
\end{figure}

\begin{figure} 
\centering\includegraphics[height=20cm,clip=true]{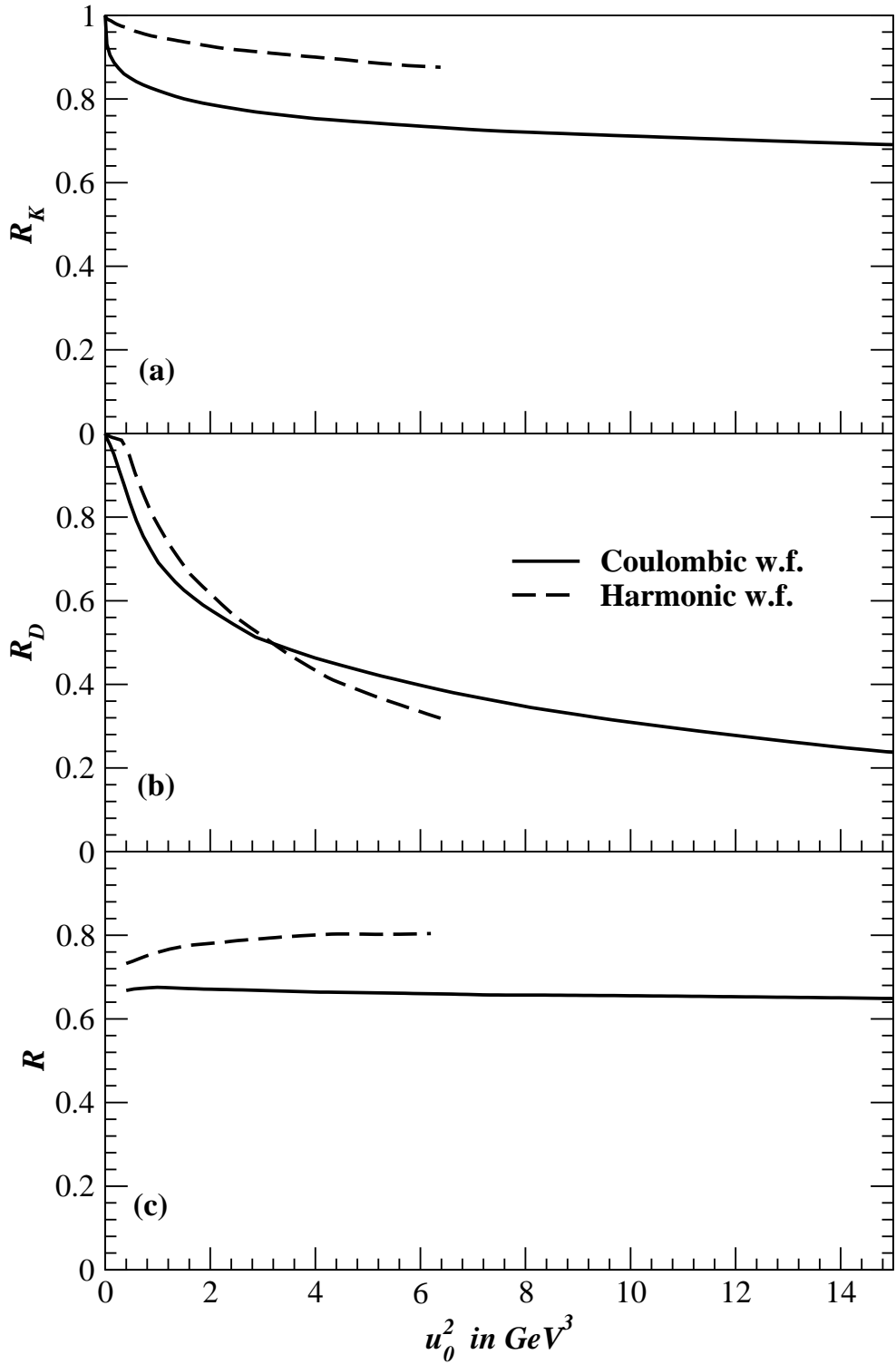}
\caption{Same as in Figs. 5 but for the Upsilon decay width. 
The bottom quark mass is taken to $M_{\Upsilon}/2$. The values of $u_0^2$ for 
the harmonic oscillator wave function have been deliberately restricted in 
order to correspond to physical ranges for the mean squared momentum between 
the quark-antiquark pair.}
\label{fig6}
\end{figure}

In Figs.~\ref{fig5} and \ref{fig6}, we plot our numerical results for the 
kinematical and dynamical (radiative) relativistic corrections, for the 
charmonium and the bottomonium ground state respectively, as a function of 
the square of the radial wave function at the origin, $u_{0}$, defined as:
\begin{equation}
\label{eq:39}
u_0 =\sqrt{4\pi} \phi(r=0)\text{ .}
\end{equation}
This is indeed the relevant variable in this case since the leading, 
non-relativistic decay width, $\Gamma_0^{NR}$, as given in Eq.~(\ref{g0}), is 
directly proportional to it. In the following numerical calculation, we 
choose $\alpha_s =0.3$ for the charmonium sector and $\alpha_s =0.15$ for the 
bottomonium sector.
 
The kinematical relativistic correction is shown in Fig.~\ref{fig5}(a) for 
the $J/\Psi$, with $m_{c}=\frac{M_{J/\Psi}}{2}$,  is the same as the one 
already calculated in Ref.~\cite{loui00} for three different phenomenological 
potentials. The results obtained here for our two kinds of test wave functions 
are very similar in shape. As expected, a larger reduction is obtained for 
the Coulombic wave function as it has larger high momentum 
components\footnote{This correction even diverges in a naive $\vec{k}^2/m^2$ 
expansion!}. The dynamical relativistic corrections are indicated in 
Fig.~\ref{fig5}(b). They are much larger, with a sharp decrease as a function 
of $u_0^2$. Again, the Coulombic wave function has a larger correction. For a 
typical value of $u_0^2$ of the order of $1\text{ GeV}^3$, the correction is 
of the order of $0.25$. This means that the well known non-relativistic 
radiative correction, $-\frac{16\alpha_s}{3\pi}$, in Eq.~(\ref{gg}) is 
reduced by a factor $1/4$. This is consistent with the first estimate of this 
correction found in Ref.~\cite{grot97}. This result has an important and 
rather nice consequence: it legitimates an expansion in $\alpha_s$ in the 
charmonium sector, since the radiative correction is now of the order of 
$10-15\%$, as compared to the $50\%$ reduction given by previous 
non-relativistic calculation with $\alpha_s=0.3$.

The total correction to the decay width, represented by the ratio $R$, is 
shown in Fig.~\ref{fig5}(c). Surprisingly enough, this ratio is rather
insensitive to the precise value of $u_0^2$ ,{\it i.e.}, to the precise value 
of the radial wave function at the origin. Indeed, the two relativistic 
corrections (kinematical and dynamical) have opposite effects on $R$ and tend 
to compensate each other, leading to an overall reduction factor varying from 
$0.6$ (for a Coulombic wave function) to $0.7$ (for an Harmonic wave 
function). This indicates that to have an agreement with the new experimental 
decay width of the $J/\Psi$ \cite{BES98}, the square of the radial wave 
function at the origin should be of the order of $0.8-1\text{ GeV}^3$. This 
is the case for several potentials, see for example 
Refs.~\cite{rich79,mart80,quig77,buch81}.  A pure Coulombic potential at 
short distance, like the Cornell potential \cite{eich78}, is however ruled 
out by our analysis, while a Coulombic potential modified according to 
asymptotic freedom  is perfectly viable \cite{rich79,buch81}.

The results for the $\Upsilon$ state are indicated on Figs.~\ref{fig6}. The
behaviour is very similar to the one observed in the charmonium sector. The 
overall correction factor is of the order of $0.65$ (for a Coulombic wave 
function) and $0.8$ (for an Harmonic wave function). The total correction 
factor is surprisingly large given the large bottom quark mass. However, the 
correction factor is much larger for a Coulombic wave function than an 
Harmonic oscillator one. This is indeed natural since for large quark mass 
the two-body wave function is concentrated to small distances. In this 
region, the high momentum components of the wave function play a major role 
in enhancing relativistic corrections.

In order to get an agreement with the experimental decay width, the square of 
the wave function at the origin needs to be of the order of 
$6$ to $8\text{ GeV}^3$. This has several interesting consequences. On the 
one hand, two potentials which are compatible with the $J/\Psi$ decay width 
(logarithmic \cite{quig77}, and power law \cite{mart80}) give a far too small 
decay width for the $\Upsilon$ \cite{quig95}. These potentials have no 
Coulombic components, and they lack in high momentum components which show up 
in the $\Upsilon$ sector. On the other hand, a pure 
Coulombic potential still gives a too large wave function at the origin. The 
two potentials of Refs.~\cite{rich79,buch81}, which correct the Coulomb 
interaction at short distances according to asymptotic freedom, are 
compatible with the experimental $\Upsilon$ decay width\footnote{The squared 
radial wave function at the origin is of the order of $6$ to 
$6.5\text{ GeV}^3$.}, as well as the $J/\Psi$ one.
  
\section{Perspectives}

We have presented a coherent formulation of the relativistic corrections to 
the leptonic decay width of heavy quarkonia. These corrections include both 
kinematical and dynamical contributions, the latter were calculated in first 
order in $\alpha_s$, but to all orders in $\vec{k}^2/m^2$.

As already found in our previous study \cite{loui00}, relativistic 
corrections are very large in the charmonium sector, and not negligible in 
the bottomonium sector.  Kinematical relativistic corrections, which are 
independent of the dynamics binding the quark-antiquark pair, lead to a large 
reduction of the leptonic decay width. Dynamical relativistic corrections, 
which correspond to relativistic corrections to the two-body bound state
itself, lead to a sizeable reduction of the standard correction
$(-\frac{16 \alpha_s}{3\pi})$ found in the non-relativistic limit. This 
result is particularly important since it shows that an expansion in
$\alpha_s$ becomes now meaningful. Indeed, the first order correction in 
$\alpha_s$ now amounts to about $10-15\%$ correction in the case of the 
$J/\Psi$, down from $50\%$ in the non relativistic limit.

While the individual relativistic corrections (kinematical and dynamical) 
depend strongly on the mean momentum squared (or equivalently on the radial 
wave function at the origin), the total relativistic correction is 
remarkably constant as a function of the square of the wave function at the 
origin. The final correction is about, $0.6$ for a Coulombic wave function 
and $0.7$ for an Harmonic wave function, relative to the non-relativistic 
zeroth order calculation (Weisskopf-Van Royen approximation).

These relativistic corrections show that realistic phenomenological
two-body potentials, such as the Richardson \cite{rich79} or the
Buchmuller-Tye \cite{buch81} potentials, are able to reproduce the decay 
width of heavy quarkonia, both in the charmonium and the bottomonium sector. 
This solves a long-standing problem in this domain, as phenomenological 
two-body potentials were able to reproduce the whole spectrum of charmonium 
and bottomonium states, but not their leptonic decay width. As already 
explained in Ref.~\cite{loui00}, the importance of relativistic corrections 
originates from the fact that the leptonic decay widths are very sensitive to 
the high momentum components of the wave function. This, because it involves 
the integral over all momentum range of the wave function and not the square 
of it, as in many other observables.

\section*{Acknowledgement}

We are grateful to V.~A.~Karmanov for fruitful discussions, and a careful 
reading of this manuscript.

\section{Appendix}

\begin{figure}
\centering\includegraphics[clip=true]{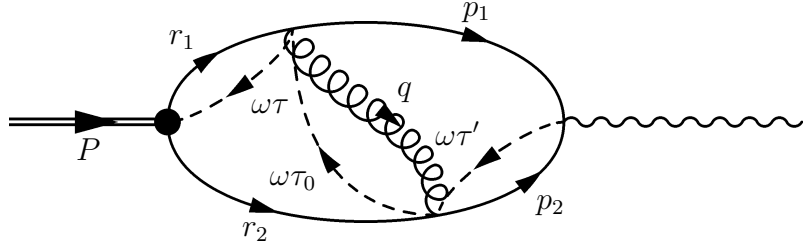}
\caption{Kinematics relevant for the first diagram contributing to the 
radiative corrections of the leptonic decay width.}
\label{fig7} 
\end{figure}

We detail in this appendix the calculation of the radiative corrections to the 
leptonic decay width. Here, we concentrate on the first diagram shown in 
Fig.~\ref{fig2}. The second one can be calculated analogously. The 
kinematical conventions are shown in Fig.~\ref{fig7}. Applying the 
diagrammatical rules given in Ref.~\cite{carb98}, we find:
\begin{multline}
\label{a:1}
{\cal M}^{\mu\rho}=-4\sqrt{\frac{m}{3}}
 \int\frac{d^3 p_1}{2\varepsilon_{p_1}(2\pi)^3} 
     \frac{d^3 r_1}{2\varepsilon_{r_1}(2\pi)^3}
     \frac{d\tau}{\tau}
     \frac{d\tau'}{\tau'}
     \frac{d\tau_0}{\tau_0} \\
 \times
  \delta(p_2^2-m^2)\delta(r_2^2-m^2)\delta(q^2)
  \Theta(\omega \sdot p_2)\Theta(\omega \sdot r_2)
  \Theta(\omega \sdot q) (s-M^2) \\
 \times
  Tr\left[\gamma^{\mu}(\SLASH{p}_1+m)
          \gamma^{\nu}(\SLASH{r}_1+m)
          \phi^\rho 
          (m-\SLASH{r}_2)\gamma_{\nu}
          (m-\SLASH{p}_2+\SLASH{\omega}\tau')
        \right]\text{ ,}
\end{multline}
with $s=(r_1+r_2)^2$. The integrations over $\tau$, $\tau'$ and $\tau_0$ can 
be done using the delta functions induced by the on-mass shell condition of 
LFD. We also use momentum conservation at vertices, including spurion momenta,
to express the various momenta. We have:
\begin{eqnarray}
\label{a:2}
\int\frac{d\tau}{\tau}\delta(r_2^2-m^2) & = & 
\frac{1}{(1-x)(s-M^2)}\text{ ,}\nonumber\\
\label{a:3}
\int\frac{d\tau'}{\tau'}\delta(p_2^2-m^2) & = & 
\frac{1}{(1-x')(s'-M^2)}\text{ ,}\nonumber\\
\label{a:4}
\int\frac{d\tau_0}{\tau_0}\delta(q^2) & = & 
\frac{1}{2\omega \sdot p(x-x')\tau_0}\text{ ,}\nonumber
\end{eqnarray}
with:
\begin{equation}
\label{a:5}
\begin{array}{c@{=}l@{\text{ }}c@{\text{ }}
              c@{=}l@{\text{ }}c@{\text{ }}
              c@{=}l}
s' & (p_1+p_2)^2 & , & 
x & \displaystyle\frac{\omega \sdot r_1}{\omega \sdot p} & , &  
x' & \displaystyle\frac{\omega \sdot p_1}{\omega \sdot p}\text{ ,} 
\\
\tau & \displaystyle\frac{s-M^2}{2\omega\sdot p} & , & 
\tau' & \displaystyle\frac{s'-M^2}{2\omega \sdot p} & , &
\tau_0 & \tau-\displaystyle\frac{m^2-r_1\sdot p_1}{\omega \sdot p (x-x')}
\text{ .}
\end{array}
\nonumber
\end{equation}
Thus, we can express $s$, $s'$ and all other invariant quantities in terms 
of the integration variables. These integration variables can be conveniently 
chosen as $x$, $x'$, $\vec{R^{\text{ }}}_T^2$, $\vec{R'}_T^2$ and 
$\vec{R^{\text{ }}}_T\sdot\vec{R'}_T$, with:
\begin{equation}
\label{a:6}
R=r_1-xp\quad\text{ and }\quad R'=p_1-x' p\text{ ,}
\end{equation}
and $\vec{R}_T$ is the transverse part of $R$ relatively to the direction of 
$\vec{\omega}$, with the consequence that $R^2=-\vec{R}_T^2$ (see the 
appendices in Ref.~\cite{carb98} for more details). The calculation of the 
five dimensional integral is done numerically using a Monte Carlo procedure.

\end{document}